\documentclass[epj]{webofc}
\usepackage[utf8]{inputenc}
\usepackage[varg]{txfonts}   
\usepackage{booktabs}
\usepackage{amsmath}
\usepackage{xcolor}
\definecolor{darkred}{rgb}{0.4,0.0,0.0}
\definecolor{darkgreen}{rgb}{0.0,0.4,0.0}
\definecolor{darkblue}{rgb}{0.0,0.0,0.4}
\usepackage[bookmarks,linktocpage,colorlinks,
    linkcolor = darkred,
    urlcolor  = darkblue,
    citecolor = darkgreen]{hyperref}
\usepackage{subfigure}
\wocname{EPJ Web of Conferences}
\woctitle{Lattice2017}
\newcommand{\Tr}{\mathrm{Tr}}
\begin{document}
%
\selectlanguage{english}
\title{%
Efficient operators for studying higher partial waves
}
\author{%
\firstname{Jia-jun} \lastname{Wu}\inst{1,2}
\fnsep\thanks{Speaker, \email{jiajun.wu@adelaide.edu.au}}
\and
\firstname{Waseem} \lastname{Kamleh}\inst{1}
\and
\firstname{Derek B.} \lastname{Leinweber}\inst{1}
 \and
\firstname{Gerrit}  \lastname{Schierholz}\inst{3}
\and
\firstname{Ross D.}  \lastname{Young}\inst{1}
\and
\firstname{James M.}  \lastname{Zanotti}\inst{1}
}
\institute{%
Special Research Centre for the Subatomic Structure of Matter (CSSM\,),
Department of Physics, University of Adelaide, Adelaide, South Australia 5005, Australia
\and
HISKP (Theory\,) and BCTP, University of Bonn, Germany
\and
Deutsches Elektronen-Synchrotron DESY, 22603 Hamburg, Germany
}
\abstract{
An extended multi-hadron operator is developed to extract the spectra
of irreducible representations in the finite volume.
The irreducible representations of the cubic group are projected
using a coordinate-space operator.
The correlation function of this operator is computationally efficient
to extract lattice spectra.
In particular, this new formulation only requires propagator
inversions from two distinct locations, at fixed physical separation.
We perform a proof-of-principle study on a $24^3 \times 48$ lattice
volume with $m_\pi\approx 900$~MeV by isolating the spectra of
$A^+_1$, $E^+$ and $T^+_2$ of the $\pi\pi$ system with isospin-2 in
the rest frame.
}
\maketitle
\section{Introduction}\label{intro}

The numerical simulation of quark and gluon fields on a finite
lattice enables a study of the hadron spectrum and strong interactions of QCD
via first principles.
Recently, lattice QCD calculations of the hadron spectrum have
achieved tremendous progress~\cite{Liu:2016kbb, Briceno:2017max}.
High partial wave states are an important topic in hadron physics,
yet it is still a challenge to study these states on the lattice.
The relevant symmetry group on a 4-dimensional lattice is reduced from
the SO(3\,) group of the infinite volume to the cubic group of a finite
volume.
As a result, the irreducible representation of the cubic group has a
nonzero overlap with infinitely many partial wave states which are the
basis states of irreducible representations of the SO(3\,) group.
Therefore the energy eigenvalues of each irreducible representation
will receive the contributions from the phase shifts in an infinite
number of partial waves~\cite{Luscher:1990ux, Luscher:1986pf}.
There have been a number of studies outlining the relations between
discrete energy levels and phase shifts, including the mixing of high
partial waves, e.g.~\cite{Luscher:1990ux, Luscher:1986pf, Bernard:2008ax,
  Gockeler:2012yj, Luu:2011ep, Wu:2015evh}.
For these reasons, the spectra of irreducible representations of the
cubic group are key to studying high partial wave states in a lattice
calculation.

The major difficulty lies in the extraction of the spectrum of each
unique irreducible representation.
In this proceedings, we introduce a novel technique for
determining the spectra of various irreducible representations of a
2-body system.
The method relies upon constructing an operator that corresponds to a
``shell'' in coordinate space, which has similarities to the ``cube''
source employed in \cite{Berkowitz:2015eaa}, where the two body operator $\psi$ is the product of two single
particle operators separated by a distance $\vec{\delta}$.
As we will show, by summing over lattice rotations, $\hat{R}$, we
construct an operator,
$\Psi_\Gamma=\sum_{\hat{R}}C(\hat{R})\,\psi_{\hat{R}}$, which projects
onto a given irreducible representation, $\Gamma$.
The correlation function of the operator $\Psi_\Gamma$ will generate a
clean spectrum of the corresponding irreducible representation
$\Gamma$.
As an exploratory exercise, we study the isospin-2 $\pi\pi$ system,
for which several alternative methods have already been explored
\cite{Dudek:2010ew, Dudek:2012gj, Beane:2011sc}.
We demonstrate that we are able to successfully determine energy
levels of the $A^+_1$, $E^+$ and $T^+_2$ representations.

\section{Formalism}\label{sec-1}

\subsection{$O_h$ group}

In the centre-of-mass (CM\,) frame in a finite volume, the basic
symmetry group of the cubic lattice for integer spin particles is the
cubic group, $O$.
The full symmetry group includes space inversions $\hat{\pi}$,
therefore, the full group should be the product of the cubic group and
spatial inversions, i.e., $O_h = O \otimes V_2$.
There are 48 elements in the $O_h$ group, and they fall into ten
different conjugacy classes.
There are ten irreducible representations of the $O_h$ group.
The character table of the $O_h$ group is shown in
Table~\ref{tab:cubic}~\cite{Gockeler:2012yj}.

\begin{table}[thb]
  \small
  \centering
  \caption{Character table of the $ O_h$ group.}
  \label{tab:cubic}
  \begin{tabular}{llllllllllll}\toprule
    $\Gamma$/Class      &     $I$       &   $8C'_3$   & $6C_4$   &  $6C'_4$  & $3C^2_4$
                                    &      $\hat{\pi}$       &   $8C'_3\times\hat{\pi}$   & $6C_4\times\hat{\pi}$   &  $6C'_4\times\hat{\pi}$  & $3C^2_4\times\hat{\pi}$ \\
   \midrule
     $A^\pm_1$                     &   $    1       $  &   $     1      $   & $     1      $  &  $      1       $  & $      1      $
                                             &   $   \pm1  $  &   $    \pm1 $   & $    \pm1 $  &  $    \pm1   $  & $     \pm1 $      \\
     $A^\pm_2$                     &   $   1        $  &   $     1      $   & $    -1      $  &  $    -1       $  & $      1      $
                                           &     $  \pm1   $  &   $     \pm1$   & $    \mp1 $  &  $    \mp1   $  & $      \pm1$      \\
     $E^\pm$                         &   $    2       $  &   $    -1      $   & $     0      $  &  $     0        $  & $      2      $
                                         &       $ \pm 2   $  &   $    \mp 1$   & $     0      $  &  $     0        $  & $    \pm  2$         \\
     $T^\pm_1$                     &   $    3       $  &   $     0      $   & $     1      $  &  $    -1        $  & $     -1      $
                                         &       $ \pm  3  $  &   $     0      $   & $  \pm   1$  &  $    \mp1   $  & $     \mp 1$       \\
     $T^\pm_2$                     &   $    3       $  &   $     0      $   & $    -1      $  &  $     1        $  & $     -1      $
                                          &      $  \pm 3  $  &  $      0      $   & $    \mp 1$  &  $     \pm1  $  & $   \mp1   $       \\
\bottomrule
  \end{tabular}
\end{table}

In the CM frame, full rotational symmetry is described by the $SO(3\,)$
group, and the familiar basis states denoted by $|lm\rangle$.
Here $l$ and $m$ are the usual quantum numbers of total and
$z$-component of angular momentum, respectively.
However, in the finite volume, angular momentum is no longer a good
quantum number since the symmetry reduces to the $O_h$ group.
The ten irreducible representations of $O_h$ in a finite volume are
then mixtures of an infinite number of angular momentum states defined
in the infinite volume.
In Table~\ref{tab:angirr}, we list the angular momentum states up to
$l=9$ contained in the ten irreducible
representations~\cite{Luu:2011ep}.

\begin{table}[thb]
  \small
  \centering
  \caption{The list of angular momenta, $l<10$, appearing in each
    irreducible representation, $\Gamma$, of $O_h$~\cite{Luu:2011ep}.}
  \label{tab:angirr}
  \begin{tabular}{llllll}\toprule
    $\Gamma$     &       $A^+_1$   &        $A^+_2$     &        $E^+$      &        $T^+_1$    &        $T^+_2$   \\
    $l$                 &     ( 0, 4, 8, ...\,) &    ( 6, ... \,)  &     ( 2, 4, 6, 8, ... \,)  &     ( 4, 6, 8, ...\,)      &     ( 2, 4, 6, 8, ... \,)  \\
    \midrule
    $\Gamma$      &       $A^-_1$   &        $A^-_2$     &        $E^-$      &        $T^-_1$    &        $T^-_2$      \\
    $l$                  &    ( 9, ... \,)  &     ( 3, 7, 9, ... \,)  &     (  5, 7, 9, ... \,)    &      (1, 3, 5, 7, 9, ...\,)   &     ( 3, 5, 7, 9, ... \,)  \\
    \bottomrule
  \end{tabular}
\end{table}

\subsection{Operator}

The standard lattice operator for the pion is :
\begin{eqnarray}
\phi(t;\vec{x}\,)\equiv\sum_a  \bar{u}^{a}(t;\vec{x}\,)\gamma_5 d^a(t;\vec{x}\,)\,,\label{eq:pionoperator1}
\end{eqnarray}
with a sum over colour indices $a$.

In coordinate space, we construct a composite operator with a
separation $\vec{\delta}$ between two pions:
 %
\begin{eqnarray}
\psi(t;\vec{x},\vec{\delta}\,)
\equiv
\phi(t;\vec{x}+\vec{\delta}/2\,) \phi(t;\vec{x}-\vec{\delta}/2\,).
\end{eqnarray}
This operator is equivalent to any other that is one of the 48
rotation operators, $\hat{R} \in O_h$ group.
We then denote the 48 pion-pion operators $\psi_{\hat{R}}$ as:
\begin{eqnarray}
\psi_{\hat{R}}(t;\vec{x},\vec{\delta}\,)=\hat{P}_{\hat{R}}\,\psi(t;\vec{x},\vec{\delta}\,)\,\hat{P}_{\hat{R}^{-1}}
=\psi(t;\hat{R}^{-1}\vec{x},\hat{R}^{-1}\vec{\delta}\,)
=\phi(t;\hat{R}^{-1}(\vec{x}+\vec{\delta}/2\,)\,)\, \phi(t;\hat{R}^{-1}(\vec{x}-\vec{\delta}/2\,)\,)\,.  \label{eq:prpsipr}
\end{eqnarray}

In this work, we maximize the symmetry available by avoiding
equivalent rotations, i.e., $\hat{R}\,\vec{\delta} \neq \vec{\delta}$.
This simply requires $\delta_x\neq\delta_y\neq\delta_z\neq0$.
While the coordinate position of each pion should be on the lattice
grid, i.e, $\hat{R}(\vec{x}\pm\vec{\delta}/2\,)\in \mathbb{Z}^3$, the
origin of the composite operator need not be on a lattice site.
One convenient choice is $\vec{\delta} = (1,\,3,\,5\,)$ and
$\vec{x}=(1/2,\,1/2,\,1/2\,)$, which keeps the shortest
$|\vec{\delta}|$.

By using the above operators, the
$|\pi^-(t,\vec{x}+\vec{\delta}/2\,)\,\pi^-(t,\vec{x}+\vec{\delta}/2\,)\rangle$
system can be generated as $\psi^\dag(t;\vec{x},\vec{\delta}\,)
|\Omega\rangle$ where $|\Omega \rangle$ is the vacuum state.
Correspondingly, 48 different states can be generated according to
Eq.(\ref{eq:prpsipr}\,) through the use of rotation operators.
For simplification, we use
$|\psi^\dag_{\hat{R}}(t;\vec{x},\vec{\delta}\,) \rangle$ to replace the
$|\pi^-(t,\hat{R}^{-1}(\vec{x}+\vec{\delta}/2\,)\,)\,\pi^-(t,\hat{R}^{-1}(\vec{x}+\vec{\delta}/2\,)\,)\rangle$
system.
These pion-pion systems should obey rotational symmetry as follows:
\begin{eqnarray}
\hat{P}_{\hat{R}}\,|\psi^\dag_{\hat{R}'}(t;\vec{x},\vec{\delta}\,) \rangle
&=&
\sum_{\hat{R}'' \in O_h}
|\psi^\dag_{\hat{R}''}(t;\vec{x},\vec{\delta}\,) \rangle
\left(\bar{B}(\hat{R}\,)\,\right)_{\hat{R}'', \hat{R}'}, \label{eq:PrB}
\end{eqnarray}
where matrix $ \left(\bar{B}(\hat{R}\,)\,\right)_{\hat{R}'', \hat{R}'} =
\delta_{\hat{R}\hat{R}',\, \hat{R}''}$, is an element of the regular
representation of the group $O_h$.
We note that we have introduced the notation where we directly use the
48 rotation operators of the $O_h$ group as indices of the matrix
$\bar{B}$.
Furthermore, here ``$\,\bar{B}\,$'' refers to the matrix
representation of $B$, e.g., the matrix of irreducible representation $\Gamma$ is written as $\bar{\Gamma}$.

The regular representation of the $O_h$ group is reducible.
Through the similarity transformation matrix, $\bar{S}$, all of these
48 matrices, $\bar{B}(\hat{R}\,)$, can be block diagonalised into a
matrix $\bar{A}$.
Each nonzero block matrix is in an irreducible representation:
\begin{eqnarray}
\bar{S}^{-1} \bar{B}(\hat{R}\,)\,\bar{S} =1 \bar{A}^\pm_1(\hat{R}\,) \oplus
1 \bar{A}^\pm_2(\hat{R}\,) \oplus 2 \bar{E}^\pm(\hat{R}\,)  \oplus 3
\bar{T}^\pm_1(\hat{R}\,)  \oplus 3 \bar{T}^\pm_2(\hat{R}\,)  \equiv
\bar{A}(\hat{R}\,)\,.
\end{eqnarray}
The number before the irreducible representation indicates the number
of copies of that representation.
Therefore, it is convenient to set the indicies of $\bar{A}$ as the
group of three numbers, $(i, \Gamma, n\,)$, where $i$ indicates the
$i$-th irreducible representation,
$\Gamma$ its name, and $n$ indicates its order.
These matrices $\bar{A}$ can be written as:
\begin{eqnarray}
\bar{A}_{i\Gamma\,n, i'\Gamma'\,n'}(\hat{R}\,) =
\delta_{i'i}\,\delta_{\Gamma'\Gamma}\,\bar{\Gamma}_{n,n'}(\hat{R}\,)\,,
\label{eq:A1}
\end{eqnarray}
Because $O_h$ is a finite group, matrices $\bar \Gamma$ can be chosen
as unitary matrices.

As shown in Eq.~(\ref{eq:PrB}\,), the matrix $\bar{B}(\hat{R}\,)$
describes the rotations between the 48 states,
$|\psi_{\hat{R}}\rangle$.
Similarly, the matrix $\bar{A}$ describes rotations between systems
classified in terms of the irreducible representations,
$|\Psi^\dag_{i,\Gamma,n}\rangle$, satisfying:
\begin{align}
  \hat{P}_{R}|\Psi^\dag_{i,\Gamma,n}\rangle =
  \sum_{i',\Gamma', n'} |\Psi^\dag_{i',\Gamma',n'}\rangle
  \left(\bar{A}(\hat{R}\,)\,\right)_{i'\Gamma'\,n', i\,\Gamma\,n}\,.
  \label{eq:PrA}
\end{align}
The transformation matrix $\bar{S}$ connects
$|\psi^\dag_{\hat{R}}\rangle$ and $|\Psi^\dag_{i, \Gamma, n}\rangle$
as follows:
\begin{align}
  |\Psi^\dag_{i, \Gamma, n}\rangle =
  \sum_{\hat{R}} |\psi^\dag_{\hat{R}}\rangle \bar{S}_{R, i\Gamma n}\,.
  \label{eq:newPhi}
\end{align}
Correspondingly, pion-pion operators that project onto irreducible
representations can be computed:
%
\begin{eqnarray}
  \Psi^\dag_{i, \Gamma, n} =
  \sum_{\hat{R}} \psi^\dag_{\hat{R}} \bar{S}_{R, i\Gamma n}\,.
  \label{eq:newPhioper}
\end{eqnarray}

\subsection{Correlation Function}
Employing the operator $\psi_R$ at the source and sink yields the
following two-point correlation function,
\begin{eqnarray}
G_{\hat{R},\hat{R}'}(t;\vec{p};\vec{x},\vec{\delta}\,) \equiv
\sum_{(\vec{y}-\vec{x}\,)\in \mathbb{Z}^3}
e^{i\vec{p}\cdot(\vec{y}-\vec{x}\,)}
\;\langle T
\left( \psi_{\hat{R}}(t;\vec{y},\vec{\delta}\,),\,
\psi^\dagger_{\hat{R}'}(0;\vec{x},\vec{\delta}\,)  \,\right) \rangle
=
G_{\hat{R}'^{-1}\hat{R},\hat{I}}(t;\vec{p};\vec{x},\vec{\delta}\,),
\label{eq:psical2}
\end{eqnarray}
while the correlation function for the two pion operator
$\Psi_{i,\,\Gamma,\,n}$ can be written as:
\begin{eqnarray}
\tilde{G}_{\Gamma\,n,\,\Gamma'\,n'}(t;\vec{p};\vec{x},\vec{\delta}\,)
=
\sum_{(\vec{y}-\vec{x}\,)\in \mathbb{Z}^3}
e^{i\vec{p}\cdot(\vec{y}-\vec{x}\,)}
\;\sum_i \langle T \left( \Psi_{i,\Gamma,n}(t;\vec{y},\vec{\delta}\,),\,
\Psi^\dagger_{{i,\Gamma',n'}}(0;\vec{x},\vec{\delta}\,) \,\right) \rangle.
\label{eq:Psicorle1}
\end{eqnarray}
Through Eq.(\ref{eq:newPhioper}\,) and the properties of group, we get:
\begin{eqnarray}
\tilde{G}_{\Gamma\,n,\,\Gamma'\,n'}(t;\vec{p};\vec{x},\vec{\delta}\,) =
\delta_{\Gamma,\,\Gamma'}\delta_{n,\,n'}\sum_{\hat{R}}\chi^{\Gamma}_{\hat{R}}
G_{\hat{R},\hat{I}}(t;\vec{p};\vec{x},\vec{\delta}\,)
\equiv
\tilde{G}_{\Gamma}(t;\vec{p};\vec{x},\vec{\delta}\,) ,
\label{eq:Psicorle2}
\end{eqnarray}
where $\chi^{\Gamma}(\hat{R}\,)$ is the character number of element
$\hat{R}$ of the $O_h$ group in the irreducible representation $\Gamma$.

\begin{figure}[htbp] \vspace{-0.cm}
\begin{center}
\includegraphics[width=0.15\columnwidth,angle=90]{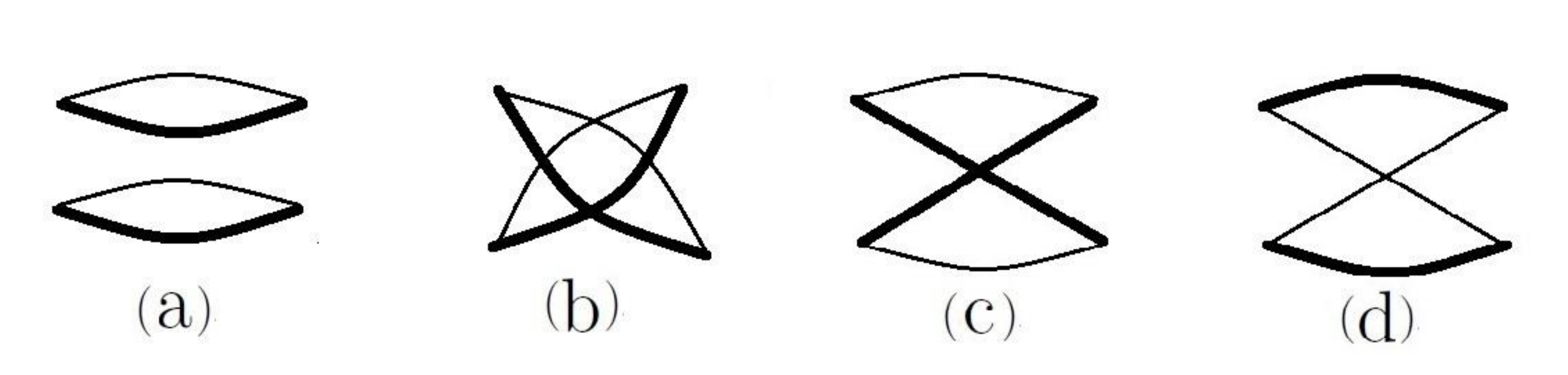}
\caption{The diagrams for the Wick contractions of $\pi\pi \to \pi\pi$
  with isospin-2. Thick and thin lines denote the $d\bar{d}$ and
  $u\bar{u}$ contractions, respectively.}
\label{fg:diagram}
\end{center}
\end{figure}

Our task is to calculate
$G_{\hat{R}^{-1},\hat{I}}(t;\vec{p};\vec{x},\vec{\delta}\,)$ as follows:
\begin{align}
\langle G_{\hat{R}^{-1},\hat{I}}(t;\vec{p}=0;\vec{x},\vec{\delta}\,)
\rangle
\hspace*{-20mm}&
\nonumber\\
&=
\sum_{(\vec{y}-\vec{x}\,)\in \mathbb{Z}^3}
\left\{\Tr\left[
S_d( \vec{y}\,^-, t;\vec{x}\,^-\,)
S^{\dag}_u(\vec{y}\,^-, t;\,\vec{x}\,^-,0\,) \right]
\Tr\left[
S_d( \vec{y}\,^+, t;\, \vec{x}\,^+,0\,)
S^{\dag}_u(\vec{y}\,^+, t;\,\vec{x}\,^+,0\,) \right] \right.\nonumber\\
&\quad\quad + \Tr\left[
S_d(\vec{y}\,^+, t;\, \vec{x}\,^-,0\,)
S^{\dag}_u(\vec{y}\,^+, t;\,\vec{x}\,^-,0\,)\right]
\Tr\left[
S_d(\vec{y}\,^-, t;\, \vec{x}\,^+,0\,)
S^{\dag}_u(\vec{y}\,^-, t;\,\vec{x}\,^+,0\,)\right] \nonumber\\
&\quad\quad - \Tr\left[
S_d( \vec{y}\,^-, t;\,\vec{x}\,^-,0\,)
S^{\dag}_u(\vec{y}\,^+, t;\,\vec{x}\,^-,0\,)
S_d(\vec{y}\,^+, t;\,\vec{x}\,^+,0 \,)
S^{\dag}_u( \vec{y}\,^-, t;\,\vec{x}\,^+,0\,) \right] \nonumber\\
& \quad\quad\left. - \Tr\left[
S_d(\vec{y}\,^+, t;\,\vec{x}\,^-,0\,)
S^{\dag}_u(\vec{y}\,^-, t;\,\vec{x}\,^-\,)
S_d(\vec{y}\,^-, t;\,\vec{x}\,^+ ,0\,)
S^{\dag}_u(\vec{y}\,^+, t;\,\vec{x}\,^+,0\,) \right]
\right\}.
\label{eq:detailcal}
\end{align}
Here,
$S_{u}(\vec{y},t;\vec{x},0\,)=T(u(\vec{y},t\,),\,\bar{u}(\vec{x},0\,)\,)$ is quark operator,
and we introduce the notation $\vec{x}\,^\pm = \vec{x} \pm \vec{\delta}/2$ and
$\vec{y}\,^\pm = \vec{y} \pm \hat{R}\vec{\delta}/2$.
Isospin symmetry $S_u \equiv S_d$  is assumed for the quark propagators.
The four terms in Eq.~(\ref{eq:detailcal}\,) correspond to the Wick contractions as shown
in Fig.\ref{fg:diagram}.

From Eq.(\ref{eq:psical2}\,) or Eq.(\ref{eq:Psicorle2}\,), two correlation functions can be calculated.
%
%
However, the first one will have a very dense spectrum, overlapping
(in principle\,) with all irreducible representations:
\begin{eqnarray}
 G_{\hat{R}^{-1},\hat{I}}(t;\vec{p};\vec{x},\vec{\delta}\,)
\sim \sum_{i} \left\{
   Z^{A^+_1}_{i}e^{-E^{A^+_1}_{i}t}
+Z^{A^+_2}_{i}e^{-E^{A^+_2}_{i}t}
+Z^{E^+_2}_{i}e^{-E^{E^+}_{i}t}
+Z^{T^+_1}_{i}e^{-E^{T^+_1}_{i}t}
+Z^{T^+_2}_{i}e^{-E^{T^+_2}_{i}t}
\right\}\,,
\end{eqnarray}
where the $\sum_i$ indicates a sum over all possible excited states
within each irreducible representation.
In fact, the ground-state energy of $A^+_1$ is dominated by an
$S$-wave state with zero momentum in the CM frame.
It will therefore have the lowest energy and hence, in principle, the
large Euclidean time behavior of this correlator will be saturated by
this eigenstate.
%
%
However, as shown in Eq.(\ref{eq:Psicorle2}\,), it is possible to
project directly onto the irreducible representation, $\Gamma$, and
hence:
\begin{eqnarray}
\tilde{G}_{\Gamma}(t;\vec{p};\vec{x},\vec{\delta}\,)  \sim \sum_{i}
   Z^{\Gamma}_{i}e^{-E^{\Gamma}_{i}t}.
\end{eqnarray}
%
In summary, we have shown how it is possible to extract the spectrum
of each unique irreducible representation by constructing the
appropriate sink-projected correlation function, $\tilde{G}$.

\subsection{L\"uscher's equation for high partial waves}
In the $\pi\pi$ system with isospin-2, only positive parity and even
angular momentum states are possible.
In our following lattice simulation, we are able to extract the ground
states of the $A_1^+$, $E^+$ and $T^+_2$ representations.
Therefore, we limit the angular momentum to $l \leq 4$.
From Ref.~\cite{Luscher:1990ux},  the formulae between partial wave
phase shifts and the energy levels can be derived.
In the case here, each irreducible representation has two partial
waves up to  $l = 4$.
Thus the eigenvalue equation will have a
$2\times 2$ matrix as follows:
\begin{eqnarray}
\det
\left( \begin{array}{cc}
  cot \delta_{l_1}(q_{\Gamma}\,) + M^{\Gamma}_{11}(q_{\Gamma}\,)
&  M^{\Gamma}_{12}(q_{\Gamma}\,)  \\
  M^{\Gamma}_{12}(q_{\Gamma}\,)
&  cot \delta_{l_2}(q_{\Gamma}\,) + M^{A_1}_{22}(q_{\Gamma}\,)
         \\
\end{array} \,\right) = 0\,, \label{eq:mom0rel1}
\end{eqnarray}
where
\begin{eqnarray}
M^{\Gamma}_{ij}(q\,)=\sum_{s} C^{\Gamma}_{i,j,s}
\frac{Z_{s0}(1;\frac{qL}{2\pi}\,)}{\sqrt{(2s+1\,)\pi^3}\left(\frac{qL}{2\pi}\,\right)^{s+1}}\,.
\label{eq:mom0relM}
\end{eqnarray}
The indices $l_1,\,l_2$ indicate which two angular momenta are
contained in the irreducible representation $\Gamma$
$q(\Gamma\,)$ is the on-shell momentum of the energy level of
irreducible representation $\Gamma$, while the function $Z_{s0}$ is
the zeta function defined in Ref.~\cite{Luscher:1990ux}.
The indices $l_1,\,l_2$, $s$ and the coefficient $ C^{\Gamma}_{i,j,s}$
are all listed in Table~\ref{tab:restphase}.

\begin{table}[thb]
  \small
  \centering
  \caption{The values of indexes $l_1$, $l_2$, $s$ and coefficients $
    C^{\Gamma}_{i,j,s}$ for each irreducible representation.}
  \label{tab:restphase}
  \begin{tabular}{llll}\toprule
    $\Gamma$ \,\,\,  &   $(l_1,\, l_2\,)$\,\,    &     $(i,\, j\,)$\,\,   &  $(s, C^{\Gamma}_{i,j,s}\,)$\,\,     \\
    \midrule
     $A^+_1$           &   (0,\,4\,)               &    $ (1,\,1\,)$       &  $ (0,\, 1\,)$ \\
                                &                            &    $ (1,\,2\,)$       &  $ (4,\, \frac{6\sqrt{21}}{7}\,)$ \\
                                &                            &    $ (2,\,2\,)$
                                &  $ (0,\, 1\,)$,\, $ (4,\, \frac{224}{143}\,)$,\, $ (6,\, \frac{80}{11}\,)$,\, $ (8,\, \frac{560}{143}\,)$\\
     $E^+$                &   (2,\,4\,)              &    $ (1,\,1\,)$       &  $ (0,\, 1\,)$,\, $ (4,\, \frac{18}{7}\,)$ \\
                                &                            &    $ (1,\,2\,)$       &  $ (4,\, -\frac{120\sqrt{3}}{77}\,)$,\,$ (6,\, -\frac{30\sqrt{3}}{11}\,)$ \\
                                &                            &    $ (2,\,2\,)$
                                &   $ (4,\, \frac{324}{1001}\,)$,\, $ (6,\, -\frac{64}{11}\,)$,\, $ (8,\, \frac{392}{143}\,)$\\
     $T^+_2$            &   (2,\,4\,)              &    $ (1,\,1\,)$       &  $ (0,\, 1\,)$,\, $ (4,\, -\frac{12}{7}\,)$ \\
                                &                            &    $ (1,\,2\,)$       &  $ (4,\, -\frac{60\sqrt{3}}{77}\,)$,\,$ (6,\, \frac{40\sqrt{3}}{11}\,)$ \\
                                &                            &    $ (2,\,2\,)$
                                &    $ (0,\, 1\,)$,\, $ (4,\, -\frac{162}{77}\,)$,\, $ (6,\, \frac{20}{11}\,)$\\
    \bottomrule
  \end{tabular}
\end{table}

\section{Results}

\subsection{Numerical calculation}

As it stands, Eq.~(\ref{eq:Psicorle2}\,) implies calculating 48
$G_{\hat{R},\hat{I}}$ correlation functions corresponding to the
different rotations $\hat{R}$.
However, for each measurement, we only require two sources for the
quark propagators at positions at $\vec{x} \pm \vec{\delta}/2$.
The 48 rotations can be summed over at the sink of the correlation
function.
This significantly reduces the computational overhead of additional
propagator inversions.

\begin{figure}[htbp] \vspace{-0.cm}
\begin{center}
\includegraphics[width=0.7\columnwidth]{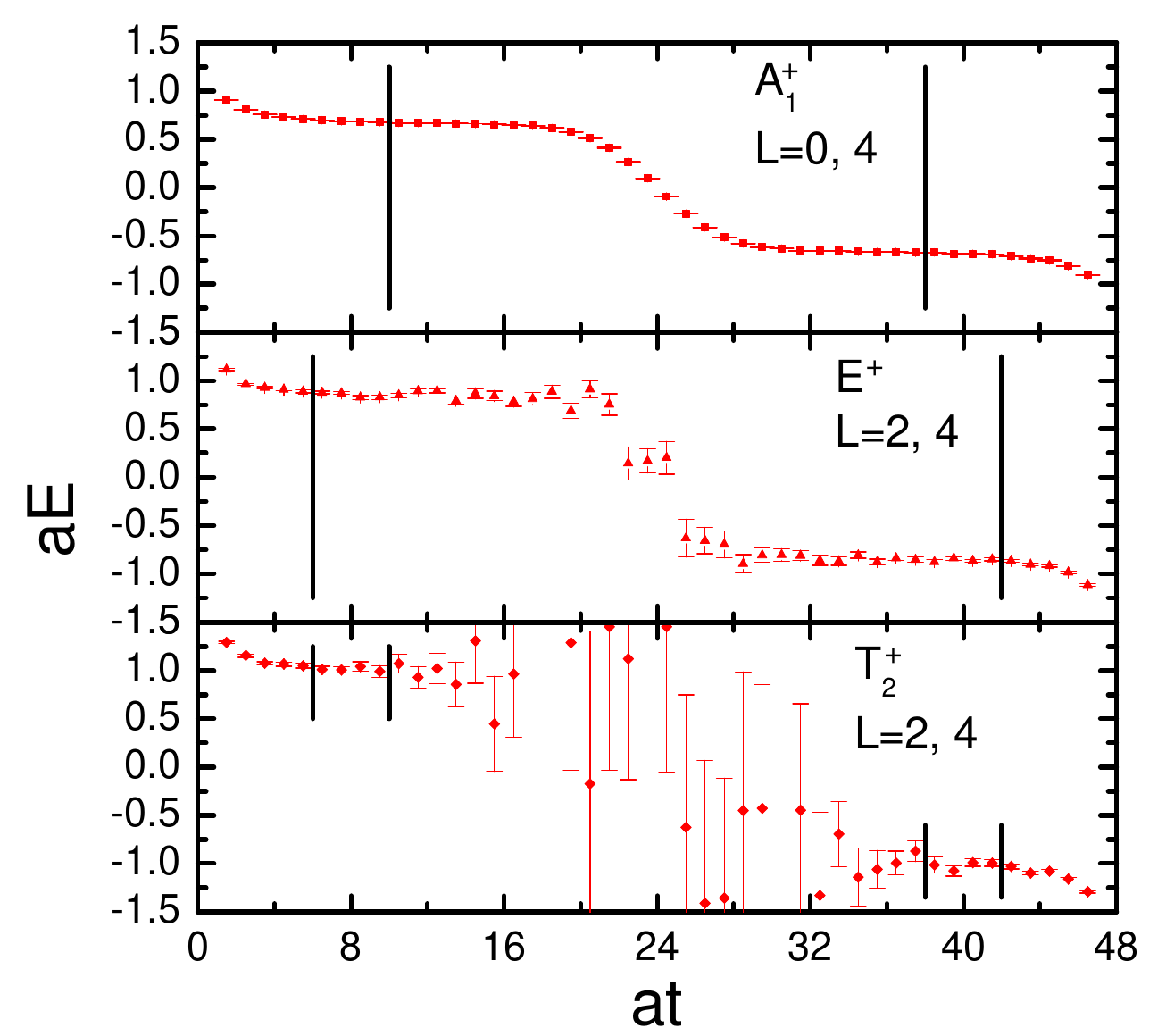}
\caption{The effective energy $aE$ of $A^+_1$, $E^+$ and $T^+_2$
  representations for the rest frame on a $24^3\times 48$ lattice with
  $a=0.071$~fm and $m_\pi\approx 900$~MeV.
  The thick black lines show the limits of the fitting windows.}
\label{fg:mom024}
\end{center}
\end{figure}

The present calculation is performed on an ensemble with 2 flavours of
dynamical ${\cal O}(a\,)$-improved Wilson fermions with
$(\beta,\,\kappa,\,V\,)=(5.29,\,0.13550,\,24^3\times 48\,)$, corresponding
to $(a,\,m_\pi=(-0.071\,{\rm fm},\, 900\,{\rm MeV}\,)$
~\cite{Bali:2012qs}.
Results are collected from 376 configurations using 16 different
source locations, with 2 inversions for each compound source.
In Fig.~\ref{fg:mom024}, the plateau of
$\log(\tilde{G}_{\Gamma}(t)/\tilde{G}_{\Gamma}(t+1)\,)$ are identified
in the $A^+_1$, $E^+$ and $T^+_2$ representations.
By choosing suitable fitting windows, the energy levels are
determined.
Our (preliminary\,) results for the corresponding on-shell momenta are:
$\frac{p_{A_1}L}{2\pi} = 0.2455(36\,)$,
$\frac{p_{E}L}{2\pi} = 1.047(13\,)$, and $\frac{p_{T_2}L}{2\pi} =
1.457(23\,)$.
As shown in Ref.~\cite{Luu:2011ep, Dudek:2012gj}, the lowest momentum
the non-interacting two-pion state in the $A^+_1$ can have is
$|\vec{p}L/2\pi|=0$, while for $E$ it is $|\vec{p}L/2\pi|=1$ and $T_2$ is
$|\vec{p}L/2\pi|=\sqrt{2}$.
Our determined momenta appear to indicate a weak repulsion in each of
these channels.
%
%
It is certainly encouraging that the ground-state energies in each of
the irreducible representations are being reliably determined.

By using this method, the momentum of each pion has not been
specified.
Only the total momentum of the system is input ---
for which we have only considered $\vec{P}=\vec{0}$ in this study.
In contrast to formulations that involve a momentum-projected hadron
at the sink \cite{Beane:2011sc,Berkowitz:2015eaa}, we use the same
operator at both source and sink.
This method then lends itself to a computationally-efficient technique
for extension to a variational analysis \cite{Blossier:2009kd}, where
the operator basis can be extended by varying $\delta$.
%
%

\subsection{Phase shifts}

\begin{figure}[htbp] \vspace{-0.cm}
\begin{center}
\includegraphics[width=0.7\columnwidth]{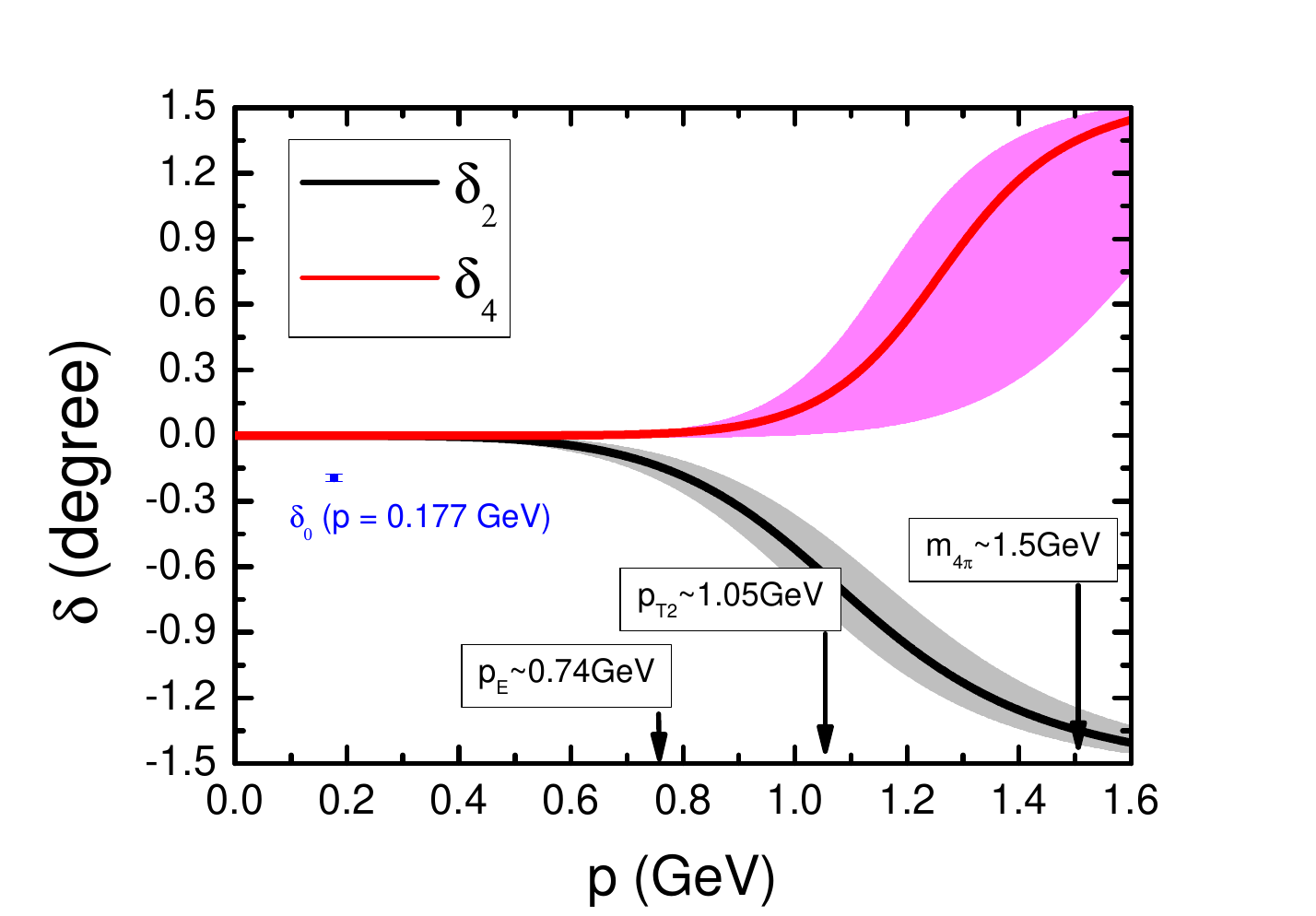}
\caption{The fitting result from the spectra in the rest frame
  with lattice size $L=24$ and spacing $a=0.071$~fm.
  The two parameters $a_2$ and $a_4$ are $-0.571 \pm 0.155$
  GeV$^{-5}$ and $0.114 \pm 0.100$ GeV$^{-9}$, respectively. The value
  for $\delta_0$($p=0.177 $GeV\,) = $-0.190\pm 0.017$.}
\label{fg:restfit}
\end{center}
\end{figure}

Equation~(\ref{eq:mom0rel1}\,) defines a relationship between phase
shifts and the finite-volume energy levels.
Since we only have one eigenvalue in each irreducible representation,
at three different on-shell momenta, we cannot solve for the phase
shifts directly.
For illustration, we parameterise the phase shifts in terms of an
effective range expansion as follows~\cite{Hoogland:1977kt,
  Cohen:1973yx, Zieminski:1974ex, Durusoy:1973aj, Dudek:2012gj}:
\begin{eqnarray}
q^{2l+1} \delta_l = -\frac{1}{ a_l } + \frac{1}{2}r_l\, q^2 \,.
 \label{eq:relationphpar1}
\end{eqnarray}
%
%
Since the S-wave only appears in the $A^+_1$ irreducible representation, we can solve for
this phase shift directly.
With the present calculation, the best we can do is then extract the
scattering length for $l=2,4$, neglecting the $r_l$ terms.
%
%
In Fig.~\ref{fg:restfit}, we show the phase shifts based on the
determination of the scattering lengths.
A strong signal for $\delta_2$ is observed.
The determination of $\delta_4$ is clearly pushing the limits of our
statistical precision. Clearly we do not anticipate attraction in this
channel, yet it is encouraging that we are on the cusp of producing a
meaningful signal for $l=4$.
Obviously, extending to a variational analysis to gain access to more
excited states, and including boosted states for a more dense spectrum
\cite{Dudek:2012gj} will dramatically improve the available
constraints.

\section{Summary}

In this paper, we introduce a new extended operator to extract the
spectra of irreducible representations.
In coordinate space, the two-particle operator projects onto an
irreducible representation by summing appropriately over a spherical
shell.
The method is computationally effecient, as it only requires two
source inversions per measurement.
Having the same operator as source and sink will allow a
straightforward implementation of a variational analysis.

For the numerical investigation in this work, we studied the isospin-2
$\pi\pi$ system in the CM frame on a $24^3\times 48$ lattice with a
lattice spacing of $a=0.071$~fm and $m_\pi\approx 900$~MeV.
The spectra of $A^+_1$, $E^+$ and $T^+_2$ irreducible representations
of a $\pi^-\pi^-$ system are successfully extracted.
The phase shifts can be calculated through the L\"uscher equation with
a toy model parameterisation.
In the future, this method can also be readily extended to boosted frames,
where the rotational symmetries on the lattice are reduced to the
little groups.

\section*{Acknowledgements}
The simulations were undertaken using supercomputing resources
provided by the Phoenix HPC service at the University of Adelaide and
at the NCI National Facility in Canberra, Australia.
The NCI resources are provided through the National Computational
Merit Allocation Scheme and the University of Adelaide Partner Share
supported by the Australian Government.
We have employed the Chroma software library~\cite{Edwards:2004sx} in
our analysis.
This investigation has been supported by the Australian Research
Council under grants FT120100821, FT100100005 and DP140103067.
GS was supported by DFG grant SCHI 179/8-1.

\bibliography{lattice2017}

\begin{thebibliography}{19}

\bibitem{Liu:2016kbb}
C.~Liu, PoS \textbf{LATTICE2016}, 006 (2017), \texttt{1612.00103}

\bibitem{Briceno:2017max}
R.A. Briceno, J.J. Dudek, R.D. Young (2017), \texttt{1706.06223}

\bibitem{Luscher:1990ux}
M.~L{\"u}scher, Nucl. Phys. \textbf{B354}, 531 (1991)

\bibitem{Luscher:1986pf}
M.~L{\"u}scher, Commun. Math. Phys. \textbf{105}, 153 (1986)

\bibitem{Bernard:2008ax}
V.~Bernard, M.~Lage, U.G. Meissner, A.~Rusetsky, JHEP \textbf{08}, 024 (2008),
  \texttt{0806.4495}

\bibitem{Gockeler:2012yj}
M.~G{\"o}ckeler et~al., Phys. Rev. \textbf{D86}, 094513 (2012),
  \texttt{1206.4141}

\bibitem{Luu:2011ep}
T.~Luu, M.J. Savage, Phys. Rev. \textbf{D83}, 114508 (2011), \texttt{1101.3347}

\bibitem{Wu:2015evh}
J.J. Wu, T.S.H. Lee, D.B. Leinweber, A.W. Thomas, R.D. Young, JPS Conf. Proc.
  \textbf{10}, 062002 (2016), \texttt{1512.02771}

\bibitem{Berkowitz:2015eaa}
E.~Berkowitz et~al., Phys. Lett. \textbf{B765}, 285 (2017), \texttt{1508.00886}

\bibitem{Dudek:2010ew}
J.J. Dudek et~al., Phys. Rev. \textbf{D83}, 071504 (2011), \texttt{1011.6352}

\bibitem{Dudek:2012gj}
J.J. Dudek, R.G. Edwards, C.E. Thomas, Phys. Rev. \textbf{D86}, 034031 (2012),
  \texttt{1203.6041}

\bibitem{Beane:2011sc}
S.R. Beane et~al. (NPLQCD), Phys. Rev. \textbf{D85}, 034505 (2012),
  \texttt{1107.5023}

\bibitem{Bali:2012qs}
G.S. Bali et~al., Nucl. Phys. \textbf{B866}, 1 (2013), \texttt{1206.7034}

\bibitem{Blossier:2009kd}
B.~Blossier et~al., JHEP \textbf{04}, 094 (2009), \texttt{0902.1265}

\bibitem{Hoogland:1977kt}
W.~Hoogland et~al., Nucl. Phys. \textbf{B126}, 109 (1977)

\bibitem{Cohen:1973yx}
D.H. Cohen, T.~Ferbel, P.~Slattery, B.~Werner, Phys. Rev. \textbf{D7}, 661
  (1973)

\bibitem{Zieminski:1974ex}
A.~Zieminski et~al., Nucl. Phys. \textbf{B69}, 502 (1974)

\bibitem{Durusoy:1973aj}
N.B. Durusoy et~al., Phys. Lett. \textbf{45B}, 517 (1973)

\bibitem{Edwards:2004sx}
R.G. Edwards, B.~Joo, Nucl. Phys. Proc. Suppl. \textbf{140}, 832 (2005),
  \texttt{hep-lat/0409003}

\end{thebibliography}

\end{document}